\documentclass[preprint,aps,floatfix]{revtex4}
\begin{document}
\preprint{IMSc/2004/05/20}

\title{ICGC-2004 Conference Overview}

\author{Ghanashyam Date}
\email{shyam@imsc.res.in}
\affiliation{The Institute of Mathematical Sciences, CIT Campus, Tharamani,
Chennai 600113}
\begin{abstract}
This is a written, expanded version of the summary talk given at the
conclusion of the ICGC-2004 held at Cochin. Brief introductory remarks 
are included to provide a slightly wider context to the theme talks.
\end{abstract}

\maketitle
\section{Introduction}
The fifth conference in the series of conferences titled `International 
Conference on Gravitation and Cosmology', ICGC-2004, was held at Cochin 
during January 5--10, 2004. It had 17 plenary talks and as a new feature
it also had 8 short talks which were more specialized than the plenary
talks but still accessible to a wider audience. There were three focus
themes: Cosmology, Gravitational Waves and Quantum Gravity. The summary 
below is a `bird's eye view' more in the spirit of: `what do I take home 
from the conference?' 

The summary is organized into sections corresponding to the theme
topics. Apart from the talks on the theme topics, there were also a few
which formed a group by themselves, not directly related to the main
themes. I have grouped them together under a somewhat lighter heading of
`Culture Talks'. Of course, these were neither lighter in their content 
nor in their importance, but relative to the more tightly focused set of
theme-talks they were more relaxing. A grouping of the talks is given in 
the table below.

\begin{table}[ht]
\centerline{Grouped List of Talks and Speakers}
\vskip 0.3cm
\begin{tabular}{|c|c|c|}
\hline 
& Testing GR & Clifford M. Will \\ 
& Collapse and Naked Singularities & Tomohiro Harada \\
`Culture' & Statistics of Short GRBs & Patrick Das Gupta \\ 
& Traversable Wormholes & Sayan Kar \\
& Kerr-Schild Space-times & Roy P. Kerr \\ 
\hline
& CMBR & Robert Crittenden \& Manoj Kaplinghat \\
& Complementary Observations & Jerry Ostriker \\
Cosmology & Cluster Surveys & Subhabrata Majumdar \\
& Weak Lensing & Bhuvnesh Jain \\ 
& Dark Energy Desperation & Robert Crittenden (Edmund Copeland) \\ 
\hline
& Detector Runs & Gabriela Gonzalez \\
& Event Rate Expectations & Vicky Kalogera \\
~~~~~Gravitational~~~~~ & Instrumentation Physics & Biplab Bhawal \\
Waves & Data Analysis & Sanjeev Dhurandhar \\
& Waveform Computations & Luc Blanchet \& Masaru Shibata \\
& Neutron Star equation of State & Frederic A. Rasio \\
\hline
& Strings & Sandip Trivedi \\
& Brane Cosmology & Misao Sasaki \\
Quantum & Loop Quantum Gravity & Jorge Pullin \\
Gravity & Loop Quantum Cosmology & Martin Bojowald \\
& Black Hole Entropy & Saurya Das \& Parthasarathi Majumdar \\
& Testing Quantum Gravity & Jorge Pullin \\
\hline
\end{tabular}
\end{table}
\section{The `Culture talks'}
Although believers don't always need a proof, it is nice to be assured
there are some who help us keep our faith after a scrutiny. {\em
Clifford Will} told us that GR continues to fare well in the traditional 
tests. With the possibility of direct detection of gravitational waves in 
the near future, further tests become conceivable. Notable among these are
tighter bounds on scalar-tensor theories and testing of speed of gravity
(or mass of the graviton). Since non-GR theories generically predict a
dipole radiation from an unsymmetrical binary system, the gravitational
wave experiments are expected to provide constraints on such theories.
The `phasing formula' can provide bounds on the $\omega$ parameter but
the real improvement is expected only after LISA. Speed of gravity
influences the form of the chirp signal from the in-spiral hence its
detection would also help put bounds on the graviton mass.

Naked singularities refuse to go away completely. {\em Tomohiro Harada} gave 
a status report of their occurrence mostly in the context of spherical collapse.
Notably, in the perfect fluid models with an equation of state of the
form $p = k \rho$, occurrence of naked singularities is {\em generic}.
Stability with respect to non-spherical perturbations, however, is not
known. Not much definitive seems to be known about non-spherical
collapse (except for Hayward's result that there is no horizon in
cylindrical collapse). Regarding the possibility of gravitational wave
emission from the vicinity of a naked singularity, it was mentioned
that it is possible and although some of the Weyl curvatures diverge at
the Cauchy horizon, the GW flux remains finite. Another interesting
aspect was the articulation of the notion of an `effective naked
singularity' i.e. region of high enough curvature, where classical GR may
breakdown, being visible from infinity. Their occurrence with non-zero
probability could be taken as a violation of censorship. Here some
quantum gravity input may be needed to decide the regime of breakdown of
classical GR.

{\em Patrick Das Gupta} filled us in on the topic of Gamma Ray Bursts. These
seem to fall into two classes demarcated by burst duration of about 2
seconds. He presented a statistical study of the sample of 156 {\em
short} GRBs belonging to the BATSE 4B catalog. One of the features that
emerges is the further division of these GRBs according to whether the
softer photons (energy less than 1 keV) arrive before the harder ones or
the other way around. The latter types form a larger class and have
further distinguishing features in terms of correlations among GRB
parameters. Some of the implications for theoretical modeling of GRBs 
were also discussed.

{\em Sayan Kar} reported on the issue of quantifying the notion of `small' 
violation of averaged energy condition and its relevance to traversable 
wormholes while {\em Roy Kerr} reminded us of the classic topic of 
`Kerr-Schild' geometries.
\section{Cosmology}
The emphasis in this theme was almost exclusively on observational
aspects of cosmology. Observations of anisotropies of CMB, first
conclusively detected by COBE and most recently determined in
unprecedented details by WMAP, are perceived as heralding the age of
`precision cosmology'. This particular relic of the Hot Big Bang is the,
essentially undisturbed, record of directly observable earliest event in
the cosmic history $(z \sim 1100)$. On the one hand, its anisotropies
are correlated with the matter and gravitational perturbations of the
FRW models which are supposed to provide seeds for the subsequent
structure formation while on the other hand its gross level isotropy
necessitates some form of an earlier inflationary era. The inflationary
era in turn provides a possibility of linking the perturbations at the
decoupling era to earlier (pre-inflation) primordial perturbations
presumably resulting from quantum fluctuations. Because of this linkage,
CMB also provides a means of (at least partially) constraining
inflationary models. The availability of power spectra of perturbations
(as inferred from the CMB anisotropies) and the currently available as
well as future surveys of structures on smaller length scales, offers a
means for studying other unknown ingredients of the universe such as
Dark matter and Dark Energy. This aspect makes the complementary
observations other than the CMB as important. For a general perspective,
see \cite{Ellis}.

{\em Robert Crittenden} surveyed the hot news from the latest star of 
observational cosmology, the WMAP. With its special features of all sky 
observation down to about $(1/2) ^0$ angular resolution together with 
determination of polarization anisotropies, it is currently the most precise
tool of observational cosmology. Among its main confirmations are:
\begin{itemize}
\item Perturbations are adiabatic ;
\item The spectrum is scale invariant ;
\item Integrated Sachs-Wolf effect (red-shifting of CMBR photons after
LSS) is seen;
\item Reionization is rather early ;
\item Gaussianity tests seem okay.
\end{itemize}
In the `not confirmed beyond reasonable doubt' category are:
\begin{itemize}
\item Presence of Sunyaev-Zeldovich effect (Compton scattering of CMBR 
photons); 
\item The suppression of power of low multipoles is not beyond doubt and
the issue of topology of the universe is also not settled yet;
\item North-South hemispherical asymmetry in the data?
\end{itemize}

{\em Manoj Kaplinghat} elaborated on the reionization information contained in
the polarization anisotropies in CMBR. He argued that apart from the `optical
depth' (epoch of reionization), CMBR contains more information. Although
the amplitude of the scalar perturbation and the optical depth are
measured in combination, detection of the so called B-mode would indicate 
presence of tensor perturbation and/or weak lensing. The tensor
perturbations have implications for both inflation and structure
formation. However at present WMAP cannot distinguish between different
models of reionization. Future probes such as Planck and James Webb
Space Telescope could give information about the reionization era.

{\em Jerry Ostriker} surveyed the current status of observational cosmology.
He emphasized that although most precise tool is CMBR observations, it
alone is not enough. Both $\Lambda$ and quintessence models fit the WMAP
data well and even the future Planck data may not be able to resolve
these models. Further observational tools such as Lyman-$\alpha$ Cloud
Surveys, Sloan Digital Sky Survey, Supernovae surveys, cluster surveys
need to be used to gather information about small scale. All this information
is needed to constrain the `concordance model' as well as inflationary models.
Over all, cold dark matter with $\Lambda$ fares well on the large scale. 
The nature of Dark Matter is as mysterious as that of Dark Energy and
theorists were challenged to come up with well motivated alternatives. 

Among the complementary set of observations, two were discussed in some
details. {\em Subhabrata Majumdar} discussed surveys of clusters of galaxies.
The galaxy clusters have a lot of structure and so are amenable to a variety
probes such as X-rays, SZ effect, Lensing etc. In contrast to the presently
available surveys of consisting about 100 clusters, the future surveys will 
have about 10000 clusters which will permit statistical methods (eg `self 
calibration') to be applied. {\em Bhuvnesh Jain} discussed the so called weak
lensing and associated `tomography' which can provide a direct handle on the 
equation of state of the Dark Energy. While lensing refers to (typically) 
multiple images of a distant source galaxy due to intervening matter 
distribution, weak lensing refers to {\em distortions} in the images 
typically at a less than 1\% level. This is deduced by statistical analysis 
and has been detected about three years back. The ellipticity of images (shear
effects) is sensitive to the intervening gravitational potential and by
doing a tomographic analysis one can get observational information on
evolution at different red-shifts. Jain also discussed uses of weak lensing
in the context of galaxy halos and Dark Matter. Within a time frame of
about 10 years, one can expect to see these two tools contributing to
our understanding of cosmology at smaller scales.

{\em Robert Crittenden} doubled-up as {\em Edmund Copeland} who could not 
attend the conference, and described the desperation with regard to models 
of Dark Energy.
Given that Dark Energy exists and there is no observational information
regarding its properties and nature, theorists can have a field day.
However the so called {\em coincidence problem}, in particular, why the 
universe seems to be accelerating just about now, puts severe constraints 
on models of Dark Energy. Generically, accommodating a large $\Lambda$
in the early universe and a small but dominant one now, leads to fine
tuning problem. He described various and some quite desperate, theoretical
attempts in modeling Dark Energy. It seems we have to await somewhat
direct observational inputs regarding the nature of Dark Energy.
\section{Gravitational Waves}
Gravitational waves is one of the qualitatively distinct implications of
general relativity. While indirect evidence for these has already been 
seen in the Taylor-Hulse binary pulsar system, their direct detection has 
remained elusive. Historically, the first efforts for a direct detection 
were by J. Weber using a room temperature resonant bar detector. Currently, 
there are three types of detectors at various stages of development/deployment
-- resonant Bar Detectors (ALLEGRO, AURIGA, EXPLORER, NAUTILUS, NIOBE), 
(Laser) Interferometric Detectors (LIGO, VIRGO, GEO600, TAMA300, ACIGA and 
the futuristic space based LISA) and resonant Spherical Detectors
(MiniGRAIL, Sfera, Graviton). Their respective inherent designs make them 
most suitable for different types of sources or frequency ranges of 
gravitational waves. As LIGO-I is currently at the science runs stage, at the 
conference, the focus was on the LIGO-I experiment. The most promising 
sources to be detected are gravitational waves from compact binary systems 
and the focus of theoretical talks was on wave forms from such sources. If 
(or perhaps one should say when) the gravitational waves are detected to the 
extent of mapping their sources, one can hope for a gravitational waves 
astronomy which will be a completely different (non-electromagnetic) window 
to the universe. A nice introduction to gravitational waves and their 
detection can be found, for instance, in \cite{SchutzRicci}. 

{\em Gabriela Gonzalez} gave a status summary of the LIGO-I experiment. It has
passed the stage of engineering runs and has had two science runs
already with the third one underway at the time of the conference. The
science runs have lasted for a duration of about 1-2 months indicating
the stability of the operation of the detector. The sensitivity has
progressively increased and is expected to reach the design goal in the
next couple of years. At present the detector can `see' out to a
distance of about 1 Mpc but it should reach out to about 3 Mpc (for
about 1.4 $M_{\odot}$ NS-NS inspirals) . With these runs, real data (output 
of the detector) is available  both to test the data analysis strategies as 
well as to provide some preliminary bounds on some of the parameters.

An essential component of gravitational wave experiments is the expectation 
of what kind of sources can be detected and at what rate. The main focus has
been inspiralling binary systems as the most promising sources of gravitational
waves. Estimation of such sources is based on the knowledge of the
properties of actual binary systems already seen by optical/radio means.
In this regards the discovery of a new, highly relativistic (double)
pulsar system, provides for a further types of populations of sources
pushing up the expected event rates. One could also take into account the 
precessional effects, relevant mostly for NS-BH and BH-BH binaries, for
detection templates. {\em Vicky Kalogera} gave details of event rate
estimations.

In a complex detector assembly such as LIGO, the physics of the detector
subsystems and their integration itself acquires a life of its own. To ensure
reliable and predictable behavior of the detector so that detector
output can be considered as `data' (signal + noise), a good deal 
of work is needed. {\em Biplab Bhawal} discussed, in quite some details, the 
complexities involved as well as various simulation checks needed. 

After being given a reliable data, comes the crucial step of extracting
a `signal' from the `noise'. Given that the potential signals are
buried at a 1\% level in the data, very sophisticated techniques are
required for signal extraction. {\em Sanjeev Dhurandhar} gave a grand tour of
types of signals of anticipated variety (chirps, periodic ... etc), from 
unknown types of sources and a variety of methods of analysis. Eventually, 
the data analysis has to identify specific kind of wave forms coming 
from specific types of sources with specific locations in the sky. Both
the role of abstract methods in data analysis as well as computational 
requirements were highlighted.

Of course one has to know what signal to extract. As mentioned earlier,
waveforms from coalescing binaries are the most promising signals. Of
the three main phases in a coalescence, inspiral-merger-ring down, the
first is amenable to analytical methods while the latter two need
numerical simulations. {\em Luc Blanchet} discussed the analytical computation
of chirps in the perturbative framework consisting of Post-Newtonian
($c^{-1}$), Post-Minkowskian ($G$), far zone ($R^{-1}$) and (source) multipole 
expansions. The non-linearities of evolution of metric perturbations,
makes the relation between the `source multipoles' and the `radiative
multipoles' non-trivial. Back reaction of the radiation on the motion of the
emitting source tends to circularise the orbits. The components of a 
binary are treated as point objects supplemented by a method to regularize 
the self field effects. The Hadamard regularization used until now appears 
incomplete at 3PN order whereas one needs to go up to 3.5 PN order for the
calculation of the chirp signals. The regularization introduces two arbitrary
parameters: $\lambda$ in the equation of motion and $\theta$ in the flux. 
The $\lambda$ parameter has been fixed by requiring matching of (new extended)
Hadamard regularization and the coordinate invariant dimensional regularization 
with minimal subtraction, however the parameter $\theta$ remains. Modulo 
this parameter, computations to 3.5 PN order are available. The PN expansion 
seems to converge well even near the Innermost Circular Orbit to numerical 
estimates of the binding energy.

{\em Masaru Shibata} summarized the status of numerical simulations of binaries
and also of collapse. The newer inputs in these are (a) BSSN evolution system 
which is stable with respect to constraint imposition, (b) hydrodynamical 
evolutions have a more physical `high resolution shock capturing' scheme 
incorporated and (c) the gauge conditions are dynamical. A minimum grid
of about $600 X 600 X 300$ is required and supercomputer capable of
dealing with this size is available. In the absence of black holes, long
term simulations are possible and quantitative runs to about 1 \%
accuracy are possible. For template computations however, better accuracy
is needed and the future looks optimistic.

{\em Frederic Rasio} discussed how gravitational waves from binaries involving
Neutron Stars can be used to obtain constraints on the NS equation of
state. The compactness ratio, M/R, can be inferred from deviations of GW
energy spectrum from point mass behavior at the end of inspiral phase.
Combining with numerical simulations of the merger phase one can
constrain the NS equation of state.
\section{Quantum Gravity}

Now we come to the quantum gravity theme of the conference. While all
other known fundamental interactions have been incorporated in a
perturbative quantum framework, gravity has resisted such attempts. Several 
researchers are involved in the enterprise of constructing {\em a} quantum 
theory of gravity, There are two {\em main} streams (represented at the 
conference) of such people: `the Unifiers' (string theorists) and `the 
Background Independents' (loopy quantum geometers). They are distinguished
by their pursuits of almost orthogonal {\em strategies}.

The string theorists aim to build a theory which includes matter (known
and unknown) and gravity in a {\em single framework}. This is sought to
be achieved by imagining a set of `elementary quanta' interacting via
mutual exchanges. This is naturally arranged by excitations of a string 
propagating in some, typically 10 dimensional, background space-times. The 
joining and splitting of strings is thought to encode the interactions among 
the excitations. Since strings can propagate in a large set of backgrounds, 
one has a large class of {\em first quantized strings theories} (i.e. a 
special class of two dimensional conformal field theories). The various 
(partially verified) {\em duality conjectures}, suggests that many of these 
are actually alternative descriptions of the ``same theory" giving the hope 
that there exists a master theory, `M-Theory'. Its `phases' are described 
by `equivalence classes' of various first quantized strings while each of 
the first quantized string is thought to be small fluctuations about a possible
M-theory `vacuum' or ground state in the sense of a local minimum. Much
effort has gone into developing such a picture \cite{Sen}. It
naturally leads to the questions: Is there {\em any} phase which
corresponds to the world as we know it so far? How do we identify such a
phase? Is there any reason for us to be in our particular phase? 

The possibility of admitting compactifications consistent with standard
model as well as various grand unified models seems to say `yes' to the
first question. Continuing in the same vein, the discovery of cosmic
acceleration with its leading explanation in terms of a positive
cosmological constant was a serious challenge to the string theorists.
The various no-go theorems made it virtually impossible to arrange a
positive cosmological constant in any (conventionally) compactified 
string theory. Recently this block has been removed and {\em Sandip
Trivedi} discussed how this is achieved, as well as its implications.

The crucial new ingredient in bypassing the no-go theorems is the so
called `compactification with fluxes' -- various form fields when
integrated over appropriate submanifolds of the compact space have
non-zero values. This permits SUSY breaking in a controlled manner to
admit de Sitter vacua. The mechanism leads to a very, very, ... very
rich ($\sim 10^{100}$ local minima) structure for the superpotential. One
possible implication suggested was that there could be several
`inflatons' as well as several epochs prior to observable inflation.
Which of these vacua is selected (and how) is unclear at present -- it
could be that special initial condition chooses our phase or that we
happen to live in our phase because that is where we can live. 

String theory suggests more than 4 space-time dimensions and admits
the so called p-dimensional Dirichlet brane solutions in which open strings 
have their ends fixed on p-dimensional `planes' with Dirichlet boundary 
conditions. By contrast, the closed strings are free to wander in the full 
10 dimensional space-times. Via the Randall-Sundrum proposals, this has
led to Brane Cosmology. {\em Misao Sasaki} discussed this now popular, string 
inspired Brane Cosmology. In particular, he discussed the idea that apart 
from the graviton, the dilaton also propagates in higher dimensions which 
can be used to drive the inflationary scenario. He argued in detail that 
such an inflation, driven by a higher dimensional scalar is indistinguishable
from the usual slow role, four dimensional inflaton driven inflation to the 
leading order of computations.

While a unified quantum theory of `everything' {\em must} include
gravity, the converse need not be true. A quantum theory of gravity
does {\em not} have to be a unified theory of {\em all interactions}. The 
loopy quantum geometers focus precisely on constructing a quantum theory 
of geometry/gravity without insisting on unification. They however insist 
on manifest {\em background independence}. They argue that a quantum theory 
of gravity/geometry should not pre-suppose any preferred background geometry
in its basic formulation. Once such a formulation is constructed together 
with coupling of presently known forms of matter (also formulated in a 
background independent manner), one may worry about possibility of a theory 
of matter with or without unification. Such an enterprise is actually 
realized and has come to be known as {\em loop quantum gravity} (LQG).
This is a non-unified but non-perturbative quantum theory of gravity.

{\em Jorge Pullin} traced the main steps in the development of LQG: 
\begin{itemize}
\item The connection formulation -- viewing Einstein gravity as a canonical
system with the phase space of a gauge theory together with the Gauss, the 
diffeomorphism and the Hamiltonian constraints;
\item Loop representation and discrete spectra of (spatial) geometrical
operators such a area, volume, length;
\item Construction of the connection representation -- availability of a
precise kinematical Hilbert space;
\item Concrete proposal for the Hamiltonian constraint including
standard model matter.
\end{itemize}

Currently, LQG is regarded as mathematically well defined theory. However 
doubts are expressed regarding whether it ``is" a quantum theory of
{\em gravity} i.e. does it have a correct semiclassical limit -- an issue
which is still unresolved.

Taking the underlying discrete structure of the spatial geometry,
suggested by LQG, and possibly also of space-time as a point of departure, 
Pullin discussed recent ideas of his and his collaborator's about ab initio 
discrete formulations of (constrained) theories. Focusing on discrete `time'
(an integer) in particular, the dynamical evolution is achieved by {\em finite}
(i.e. non-infinitesimal) canonical transformations. For theories with 
constraints, preservation of constraints can be used to solve for the 
Lagrange multipliers and get a constraint-free theory which can be quantized
in the usual manner. He argued that the twin problems associated with a 
Hamiltonian constraint namely (a) clock variable as a Dirac observable will 
have a constant value and hence cease to be useful as a clock (``Problem of
Time") and (b) solutions of the Hamiltonian constraint being generically
distributional need a new physical inner product to be defined, both disappear
once the continuum theory with constraint is replaced by a discrete, 
constraint-free theory. In the context of quantum cosmological models, one 
can also `avoid singularity' essentially because discritizations which 
encounter the singularity form a `set of measure zero' in the space of all 
possible discritization. (This is very different from the singularity avoidance
in Loop Quantum Cosmology). The proposed method can also `solve' the 
information loss problem by having a {\em non-unitary} evolution (with respect
to a clock variable distinct from the discrete label `time' with respect to 
which the evolution is unitary). This approach has been demonstrated to work
for the Yang-Mills and BF theories as well as for quantum cosmological models.

Loop Quantum Cosmology (LQC), an area pioneered by Martin Bojowald, was
surveyed by {\em Bojowald} himself. By systematically carrying out the symmetry
reduction of LQG one arrives at LQC. This can also be construed as carrying
out a `loopy' quantization (distinct from the usual Wheeler-DeWitt 
quantization) of the classically symmetry reduced cosmological models. In 
this quantization, the discrete geometry revealed by LQG survives the symmetry
reduction in a subtle technical sense -- eigenvalues of the triad operator 
take all real values but each has a corresponding (normalizable) eigenvector.
Nevertheless, adaptation of the Thiemann procedures lead to a difference 
equation for selecting the solutions of the Hamiltonian constraint and 
also to definitions of inverse triad operators with bounded spectra. On the 
one hand this allows one to `evolve' through the classically indicated 
singularity (at zero triad eigenvalues) and on the other hand it permits
only a bounded growth of spatial curvatures and matter densities even as the 
singularity is approached. 

Among the applications of LQC are:
\begin{itemize}
\item For the isotropic models with scalar matter, one can get inflation
with graceful exit;
\item For the closed isotropic models, one can see both the freezing of
the matter field and its subsequent bounce as well as avoidance of the
big crunch singularity;
\item It is possible to have a running spectral index and a suppression
of power at low multipoles in CMBR;
\item The singularity avoidance continues to hold even for more general
diagonal, homogeneous cosmologies;
\item The anisotropic, vacuum Bianchi-IX evolution is non-chaotic and in
conjunction with the BKL conjecture suggests a simpler picture of
approach to the general singularity.
\end{itemize}

One point worth emphasizing is that LQC is being developed as a simpler
setting to unravel LQG and is not being arranged to address specific
phenomenological issues of cosmology. That it naturally admits
phenomenologically viable mechanisms is an attractive but some what incidental
feature.

Is there any way of `testing' a quantum theory of gravity?

At a theoretical level, a microscopic explanation of the black hole
entropy has been a `traditional' testing ground for QG. The test
involves reproducing the Bekenstein-Hawking formula for entropy in terms
of the horizon area. {\em Saurya Das} reviewed this entropy test for various
approaches such as Horizon CFT, Isolated Horizons in LQG, String theory,
AdS/CFT etc. Most QG candidates reproduce the entropy formula in the
leading behavior. Strings get it for extremal or near-extremal black
holes while LQG gets it for any isolated horizon modulo the value of the 
Barbero-Immirzi parameter to be fixed by normalizing to any one black hole. 
There are however logarithmic corrections at the next leading order. These 
seem to be different for different approaches and as such could be considered
discriminatory. He also surveyed efforts to understand the Hawking radiation 
in various approaches as well as discussed the information loss problem. 

{\em Parthasarathi Majumdar} discussed the micro-canonical and the canonical
entropy for general relativistic systems with a boundary. Assuming only
the area spectrum from LQG, area law for the micro-canonical entropy and
a power law relation between the energy and the area, he argued that
leading thermal fluctuation corrections to the canonical entropy are also
logarithmic with a universal coefficient. The universal form for
canonical entropy is independent of the index of the power law relation
between energy and area.

Works regarding testing QG on a more direct observational level was
reviewed by {\em Jorge Pullin}. The observation that although tiny, QG effects
could be observable due to cumulative enhancement, forms a basis for
hopes for manifestation of QG. Effects on the propagation of light and
neutrinos as well as modified dispersion relation with attendant Lorentz
invariance violations have been investigated. The hope is that 
Gamma Ray Bursts signals could contain a signature of these effects. These 
ideas are however preliminary with a rather weak theoretical 
basis/understanding. 

\section{Concluding Remarks}

In summary, the conference had a very strong emphasis on observational
aspects emphasizing applications of GR to an understanding of the cosmos 
at ever more detailed level. By the next ICGC, one may hope to get
some glimpse of the universe through the gravitational wave window. On 
the quantum gravity front, Strings seem to `suffer' from embarrassing 
riches of De Sitter Vacua ($\sim 10^{100}$) while loops have still to 
make a reliable contact with low energy (or semiclassical) limit. 

The summary won't be complete without mentioning the delightful
`Cosmos in Cartoons' by Vishu ( C. V. Vishveshwara ). These just have to 
be seen and enjoyed.

{\em Acknowledgements:}

I am grateful to Bala Iyer, Sanjeev Dhurandhar and Martin Bojowald for 
correcting my errors and other helpful remarks. I would also like to take 
this opportunity to thank the other members of the SOC and in particular
Bala Iyer, for `encouraging' me to be more than attentive during the
conference.

\end{document}